\documentclass[prd,nofootinbib,showpacs,preprint,aps]{revtex4-1}
\usepackage{subfigure}
\usepackage{amsmath}
\usepackage{amsfonts}
\usepackage{amssymb}
\usepackage[dvips]{graphicx}

\usepackage[applemac]{inputenc}
\usepackage[T1]{fontenc}
\usepackage{amssymb}
\usepackage{amsfonts}
\usepackage{amsfonts} 
\usepackage{graphicx} 

\usepackage{bbm}
\def\beq{\begin{equation}}
\def\eeq{\end{equation}}
\def\bsp{\begin{split}}
\def\esp{\end{split}}
\def\bea{\begin{eqnarray}}
\def\eea{\end{eqnarray}}
\def\ba{\begin{array}}
\def\ea{\end{array}}

\def\haf#1{{{#1}\over 2}}

\def\lb{\left(}
\def\rb{\right)}

\def\l.{\left.}
\def\r.{\right.}

\def\part{\partial}

\def\m{{\mu}}
\def\n{{\nu}}

\begin{document}
\preprint{UdeM-GPP-TH-18-262}
%\preprint{arXiv:17xx.xxx}
\title{New sources of gravitational wave signals:\\ The black hole graviton laser}
\author{Éric Dupuis}
\email{eric.dupuis.1@umontreal.ca}
\author{M. B. Paranjape} 
\email{paranj@lps.umontreal.ca}
\affiliation{Groupe de physique des particules, D\'epartement de physique,
Universit\'e de Montr\'eal,
C.P. 6128, succ. centre-ville, Montr\'eal, 
Qu\'ebec, Canada, H3C 3J7 }
\begin{abstract}\baselineskip=18pt
A graviton laser works, in principle, by the stimulated emission of coherent gravitons from a lasing medium.   For significant amplification, we must have a very long path length and/or very high densities.  Black holes and the existence of weakly interacting sub-eV dark matter particles (WISPs) solve both of these obstacles.  Orbiting trajectories for massless particles around black holes are well understood \cite{mtw} and  allow for arbitrarily long graviton path lengths.    Superradiance from Kerr black holes of WISPs can provide the sufficiently high density \cite{ABH}.   This suggests that black holes can act as efficient graviton lasers.  Thus directed graviton laser beams  have been emitted since the beginning of the universe and give rise to new sources of gravitational wave signals.  To be in the path of particularly harmfully amplified graviton death rays  will not be pleasant.

\vskip 1in
\leftline{“Essay written for the Gravity Research Foundation 2018 Awards for Essays on Gravitation.”}
\leftline{\today}
\leftline{Corresponding author: M. B. Paranjape, paranj@lps.umontreal.ca}
\end{abstract}
\maketitle
\section{Introduction}     
Black holes admit quantum mechanical bound states of ultra light particles such as axions, other dark matter candidates that are weakly interacting sub-eV particles (WISPs), \cite{WISP}, or neutrinos.  These states can undergo quantum transitions absorbing or emitting gravitons.  We are interested in considering this quantum mechanical system acting as a lasing medium for gravitons.  Graviton wave functions in the presence of black holes, in principle can correspond to gravitons that orbit the black hole arbitrarily many times before finally escaping to infinity \cite{orbitingADMY}.  In the particle picture these trajectories correspond to massless unstable orbits,  for a simple Schwarzschild black hole at $r=3GM$.  This allows for the possibility that the graviton can pass through the lasing medium for an arbitrarily large distance or equivalently, for an arbitrarily large amount of time, without the need for reflecting mirrors as is usually the case for photons in a laser.   Quantum mechanically, such graviton trajectories correspond to graviton-black hole scattering states which exhibit an arbitrarily large deflection angle or equivalently, time delay, before they emerge at infinity.   The spontaneous emission of such graviton states coupled with their subsequent stimulated emission as the graviton circles  around the black hole through the lasing medium, could give rise to significant amplification.  The observation of intense beams of coherent gravitons would then be an indirect confirmation of the existence of  ultra light massive particles.  As the coupling of gravitons to matter is so small,  very long trajectories would be required to obtain significant amplification.  It might take all the time since the beginning of the universe for the graviton to trace them out, so far this is approximately $4\times 10^{17}seconds$.  As we have not seen, to date, any particularly harmful amplified graviton rays, we can be assured that the gains have not been sufficiently large.  However, no matter what the gain, eventually, there will pass enough time for significant amplification.  

\section{Quantum gravitons}
In the lowest approximation, gravitons in the presence of a black hole correspond to the non-interacting excitations of the quantized, linearized Einstein equations in the black hole background.  These equations describe the dynamics of a massless spin 2 field moving on a Schwarzschild background geometry.  The quantization is relatively straightforward.  The quantum gravitational field corresponds to the tensor $h_{\mu\nu}$ which satisfies the equation of motion
\beq
D_\sigma D^\sigma h_{\mu\nu}(x)=0\label{eom}
\eeq
where $D_\sigma$ is the covariant derivative in the Schwarzschild background.   The graviton field can be expanded in any complete set of orthonormal wave functions $\psi_n(x)$ that satisfy the equations of motion,
\beq
h_{\mu\nu}(x)=\sum_n \left(\psi_n(x) \epsilon_{\mu\nu}(n)A( n)+\psi_n^*(x)\epsilon^*_{\mu\nu}(n)A^\dagger( n)\right)
\eeq
where $\epsilon_{\mu\nu}(n)$ is the corresponding polarization tensor  and the operators $A^\dagger(n)$  and $A(n)$  satisfy the simple algebra of annihilation and creation operators
$
\left[A(n),A^\dagger(n')\right]=\delta_{nn'}
$.
These operators respectively create and destroy one  graviton in the state $\psi_n(x)$.   Normally, plane wave states are used for the wave functions $\psi_n(x)$ however  we will be interested in a complete, orthonormal set of wave packet wave functions.   In fact we will be interested in the limit that the wave packet is localized in the radial direction at exactly $r=3GM$, while it will be a normal, wave packet in the other two independent directions.  In this way, the wave packet will have a non zero probability to orbit the black hole arbitrarily many times.  

We will perform our calculation in the approximation that the gravitational field of the black hole must be taken into account only in the way that it bends the trajectories of the gravitons.  We furthermore will work in the weak (gravitational) field limit.   Indeed, the acceleration exerted at infinity to keep an object just at the horizon is $1/4GM$, which evidently can be arbitrarily small.  
\section{Lasing medium}
The existence ultra-light WISPs with masses in the range of $10^{-11}$ to $10^{-20}$ eV implies that Kerr black holes are susceptible to the phenomenon of superradiance, \cite{ABH,AD}.  Superradiance corresponds to the process by which the WISP particles are radiated out of the Kerr black hole in copious quantities reducing its angular momentum.  This corresponds to the Penrose process \cite{PP,Penrose-process} for extracting energy and angular momentum out of a Kerr black hole, however, it is more simply understood as a form of Cerenkov radiation \cite{AD}.   The final state will be a Schwarzschild black hole surrounded by a bath of WISPs in non-zero angular momentum states, in principle, of enormous density.  Thus, the existence of ultra-light WISPs can give rise to the lasing medium.  The WISPs will be created in a population inverted state, the ground state with $l=0$ will be empty, while the states $l=1,2,\cdots$ will be highly populated.  

The Newtonian hydrogen atom of the WISP-black hole system is controlled by the effective fine structure constant
$
\alpha=GMm
$
and the bound state energies have frequencies given by
$
\omega\approx m\lb 1-\frac{\alpha^2}{2n^2}\rb\frac{c^2}{\hbar}
$
where $n$ is the principal quantum number.  The states are additionally characterized by the orbital quantum number $l_\theta$ and the magnetic quantum number $m_\varphi$.   The condition for superradiance is given by
$
\frac{\omega}{m_\varphi}<\Omega_{BH}
$.  Analysis of the superradiance in \cite{AD,ABH} yields the formula for the (maximum) occupation number of a level with magnetic quantum number $m_\varphi$
\beq
N_{\rm max}=\frac{GM^2}{\hbar c }\frac{\Delta a_*}{m_\varphi}\sim10^{76}\lb\frac{\Delta a_*}{.1}\rb\lb\frac{M}{10M_\odot}\rb^2\label{on}
\eeq
where $a_*$ is the spin of the black hole (which is, by definition, in magnitude less than one) and $M_\odot$ is the mass of the Sun and $m_\varphi$ is taken to be of order 1.  The occupied levels all have non zero angular momentum, hence they can undergo a spontaneous transition to the ground state and emit a graviton in a state that corresponds to a graviton moving arbitrarily close to the classical, unstable orbit.  This graviton can then stimulate the emission of coherent gravitons, which, given enough time, will give rise to significant amplification.  \footnote{The authors in \cite{ABH} did consider transitions between the WISP energy levels giving rise to graviton emission, however they did not consider the process of stimulated emission and the subsequent amplification.}

\section{Interaction Hamiltonian}
To compute the amplitude for stimulated emission we must first determine the interaction Hamiltonian.  We must understand the way in which the WISPs interact with the gravitational field of the graviton which is propagating on top of the background Schwarzschild geometry.  
In the weak field linear approximation, the combined metric is
\beq
d\tau^2=\left(1-\frac{2GM}{r}\right)dt^2-\left(1+\frac{2GM}{r}\right)(\vec x\cdot \vec dx)^2/r^2-r^2d\Omega^2+h_{\m\n}dx^\m dx^\n.
\eeq

The classical dynamics of the WISPs is governed by the geodesic equation
\beq
\frac{d^2x^i}{d\tau^2}+\Gamma^i_{\mu\nu}\frac{dx^\mu}{d\tau}\frac{dx^\nu}{d\tau}=0
\eeq
where $x^\m$ are the coordinates of the WISP, and canonical quantization will generate the quantum theory, \cite{GravLaser}.  We use the formalism explained by Bertschinger in \cite{bertschinger}, decomposing the gravitational field as
$
h_{00}=-2\phi,\quad h_{0i}=w_i,\quad h_{ij}=-2\psi\delta_{ij}+2s_{ij}
$
where $\delta^{ij}s_{ij}=0$.  We have $w_i=0$, $\phi=GM/r$, $\psi=GM/3r$ and $s_{ij}=s^{M}_{ij}+s^{\approx}_{ij}$ with $s^{M}_{ij}=(-GM/r^3)(x_ix_j-(r^2/3)\delta_{ij})$ and  $s^{\approx}_{ij}=\haf 1h_{ij}$ where we use the symbol $\approx$ for the gravitational wave.  We assume the gravitational wave is transverse and trace free.

The Hamiltonian for the system then generally is given by
\beq
H(x^i,\pi_j)=(1+\phi)E(p_j)\,\,{\rm where}\,\, E(p_j)=(\delta^{ij}p_ip_j+m^2)^{1/2}
\eeq
where $m$ is the WISP mass, and with $p_i=(1+\psi)\pi_i-(\delta^{ij}\pi_i\pi_j+m^2)^{1/2}w_i-s^j_{\,\,\, i}\pi_j$ which in our case becomes just $p_i=(1+\psi)\pi_i-s^j_{\,\,\, i}\pi_j$.  As $\pi_i$ and $p_i$ differ only by first order terms, we get the Hamiltonian, expanding to first order in the gravitational field and to second order in the canonical momentum 
\beq
H(x^i,\pi_j)=m +\frac{|\vec\pi|^2}{2m }+\frac{(\psi\delta^{ij}-s^{ij})\pi_i\pi_j}{m }+\phi (m +\frac{|\vec\pi|^2}{2m }).
\eeq
From this expression we extract our basic Hamiltonian of the WISP interacting with the gravitational field of the black hole
\beq
H_0(x^i,\pi_j)=\frac{|\pi|^2}{2m }+m \phi\label{hfree}
\eeq
subtracting off the rest mass, and the perturbation, which we split into two terms
\bea
H^M(x^i,\pi_j)&=& \frac{\phi|\vec\pi|^2}{2m }+\frac{(\psi\delta^{ij}-s^{M\,ij})\pi _i\pi_j}{m }\label{hearth}\\
H^\approx(x^i,\pi_j)&=&\frac{-h^{ ij}\pi _i\pi_j}{2m }.\label{hwave}
\eea

The Schrödinger equation resulting from Eqn. \eqref{hfree} is that of the Newtonian hydrogen atom
\beq
\left(\frac{-\hbar^2}{2m }\nabla^2-\frac{GMm}{r}\right)\phi(x)=E\phi(x).\label{schrodinger}
\eeq
the perturbation from Eqn.\eqref{hearth} induces some small static changes in the energies and wavefunctions of $H_0(x^i,\pi_j)$, which cannot give rise to any transitions, spontaneous or stimulated, between levels.  The time dependent perturbation that is relevant, from Eqn.\eqref{hwave}  (replacing $h_{\m\n}\rightarrow \frac{\sqrt{G}}{c^2}h_{\m\n}$ which is the canonically normalized metric perturbation) is given by
\beq
H^\approx(x^i,\pi_j)=- \frac{\sqrt{G}}{c^2}\frac{h^{ij}\pi_i\pi_j}{m }.
\eeq

\section{Amplitude for spontaneous or stimulated emission and cross section}
The amplitude for spontaneous or stimulated emission of a graviton in mode $\psi_n(x)$  is proportional to the matrix element  $\langle \phi_{f};(N+1,\psi_n)|H^\approx(x^i,\pi_j)|\phi_{i};(N,\psi_n)\rangle$ where a state $|\phi;(N,\psi_n)\rangle$ corresponds to the WISP in Newtonian hydrogen atomic state $\phi$ and $N$ gravitons in wave packet state $\psi_n$ (there could be  many other occupied states in the system, that are just spectators in the transition, which we have not listed in the ket).  Although we insisted on expanding the graviton states in terms of wave packets, it is clear that the wave packets themselves can be expanded in terms of plane waves.  For plane wave graviton states the golden rule applies and gives the rate
\beq
\Gamma_{\phi_i\to\phi_n}=\frac{2\pi}{\hbar}\delta\lb E_f-E_i-\hbar\omega\rb \langle \phi_{f};(N+1,\vec k,\omega)|H^\approx(x^i,\pi_j)|\phi_{i};(N,\vec k,\omega)\rangle
\eeq 
Modifying a calculation in Baym, \cite{gordon1990lectures} and following closely the calculation done in \cite{GravLaser} it is straightforward to obtain the total absorption cross section.  We find, surprisingly, that  the cross section is proportional to the Planck area
\beq
\sigma\sim \frac{\hbar G}{c^3}\beta
\eeq
where $\beta$ is a pure number that is parametrically of order 1, that depends on the specific geometry of the quantum system.    For the case of  quantum bouncers (ultra cold neutrons, \cite{GravLaser}), $\beta$ is a function of the difference of the zeros of the Airy function, for the case here with particles bound in the Schrodinger levels of essentially a Newtonian, hydrogen atom, $\beta$ is a function of a difference of the principle quantum numbers involved in the transition. 

Interestingly, the total absorption cross section on a source that is gravitationally defined, is universal in that it is independent of the masses of the particles involved.  The cross section is proportional to the Planck area.  We conjecture that this universality is an expression of the equivalence principle.  
\section{Gain}
The gain per unit length is given by the product of the cross section with the density $\kappa=\sigma\times\eta$.  The density of the lasing medium is approximately the occupation number, obtained from Eqn.\eqref{on}, divided by the volume of the black hole $\lb\frac{4\pi}{3}\rb\lb\frac{2GM}{c^2}\rb^3$  thus $\eta\sim \lb\frac{GM^2}{\hbar c }\rb\lb\frac{\Delta a_*}{m_\varphi}\rb\lb\frac{3}{4\pi}\rb\lb\frac{c^2}{2GM}\rb^3$, which has units of $1/meters^3$.  This gives rise to a gain 
\beq
\kappa\sim \frac{\hbar G}{c^3}\frac{GM^2}{\hbar c }\frac{\Delta a_*}{m_\varphi}\frac{3}{4\pi}\lb\frac{c^2}{2GM}\rb^3=\frac{c^2}{8GM}\frac{\Delta a_*}{m_\varphi}\frac{3}{4\pi}\approx\frac{c^2}{GM}\times .003=\frac{.003}{R_S}
\eeq 
which has units of ${1}/{meter}$, and where $R_S$ is the Schwarzschild radius and assuming $\Delta a_*\approx .1$ and $m_\varphi\sim 1$.  
\section{Discussion}
For a solar mass black hole, the Schwarzschild radius is about 3000 meters, which gives a gain of $\kappa_\odot\sim 10^{-6}/meter$.   For a super massive blackhole of $10^{10}$ solar masses, the gain is 
$\kappa_{SMBH}\sim 10^{-16}/meter$.  For significant amplification $A_F$, the path length $L$ must be long enough so that $A_F=\kappa\times L\gg 1$.  Thus $L\gg 10^6\to 10^{16}meters$, which corresponds to time scales $T$ in the range of $T\gtrsim 10^{-3}\to 10^{7}seconds$.  Thus we predict that black holes with enormous WISP clouds should have emitted coherent bursts of gravitational radiation in time scales of a few milliseconds to a few years.  Observations of these bursts would be important to solidify our understanding of black holes, gravitons and WISPs. 

Of course, the calculations leading to the enormous occupation numbers of WISPs could be greatly different, in reality.  The calculation of the absorption cross section, however, is not susceptible to much uncertainty.  Therefore, it could be that the gains are, in reality, much smaller.  If the gain is of the order of $\kappa\sim 10^{-25}/meter$ then there has only been enough time since the beginning of the universe so that the first burst of coherent gravitons would only be appearing now!  Furthermore, the time at which the coherent graviton beams appear depends crucially on which mode is amplified.  An emitted mode could be so close to the direction of the exact unstable graviton orbit that it could still to this day, be getting amplified by the lasing medium, actually leaving the black hole at some time in the future.  Particularly lethal death rays could be emitted at any time forward.  In that way, black holes with large WISP clouds in excited states are ticking time bombs.  As there has been no observable evidence of such destructive rays, we can safely assume that the gains are in fact  substantially smaller than $10^{-25}/meter$!

The unique properties of laser gravitational radiation  are  monochromaticity, directionality, coherence and  high intensity.  Observation of such signals at gravitational wave observatories would be a stunning verification of properties of black holes and of the WISP contribution to dark matter.  
\section{Acknowledgments} We thank NSERC of Canada for financial support.  We thank Jorge Gamboa and the Physics Department of the Universidad de Santiago,  Santiago, Chile, for hospitality, where this essay was written up. 

\bibliographystyle{apsrev}
\bibliography{black-hole-graviton-laser}

\end{document}